\documentstyle[11pt,aas2pp4,psfig]{article}

\def\hii{\ion{H}{2}}

\def\heii{\ion{He}{2}}

\def\niii{\ion{N}{3}}
\def\ciii{\ion{C}{3}}
\def\civ{\ion{C}{4}}
\def\ariv{\ion{Ar}{4}}
\def\feiii{\ion{Fe}{3}}
\def\oi{\ion{O}{1}}
\def\oiii{\ion{O}{3}}
\def\siii{\ion{S}{3}}
\def\halpha{\ifmmode {\rm H{\alpha}} \else $\rm H{\alpha}$\fi}
\def\hbeta{\ifmmode {\rm H{\beta}} \else $\rm H{\beta}$\fi}
\def\masl{\ifmmode  {\rm M_{\sun}yr^{-1}} \else ${\rm M_{\sun}yr^{-1}}$\fi}

\def\mdot{\ifmmode  \dot{M} \else $\dot{M}$\fi}
\def\msun{\ifmmode M_{\odot} \else $M_{\odot}$\fi}
\def\vinf{\ifmmode v_{\infty} \else $v_{\infty}$\fi}
\def\teff{\ifmmode T_{\rm eff} \else $T_{\rm eff}$\fi}
\def\logg{\ifmmode \log g \else $\log g$\fi}
\def\loggeff{\ifmmode \log g_{\rm eff} \else $\log g_{\rm eff}$\fi}
\def\rstar{\ifmmode R_{\star} \else $R_{\star}$\fi}
\def\lstar{\ifmmode L_{\star} \else $L_{\star}$\fi}
\def\mstar{\ifmmode M_{\star} \else $M_{\star}$\fi}
\def\rsun{\ifmmode R_{\odot} \else $R_{\odot}$\fi}
\def\lsun{\ifmmode L_{\odot} \else $L_{\odot}$\fi}
\def\12c16o{$^{12}{\rm C}\left(\alpha,\gamma\right)^{16}{\rm O}$}
\def\kms{\ifmmode {\rm km \;s^{-1}} \else $\rm km \;s^{-1}$\fi}

\received{2 November 1996}
\accepted{13 March 1997}

\begin{document}

\title{Detection of Wolf-Rayet stars of WN and WC subtype in 
Super Star Clusters of NGC 5253\footnote{Based on 
observations obtained at the European Southern Observatory, Chile}}

\author{Daniel Schaerer\altaffilmark{2,3}}
\altaffiltext{2}{Space Telescope Science Institute, 3700 San Martin Drive, 
Baltimore, MD 21218; schaerer@stsci.edu}
\altaffiltext{3}{Observatoire de Gen\`eve, CH-1290 Sauverny, Switzerland}

 \author{Thierry Contini\altaffilmark{4,5}}
\altaffiltext{4}{UMR 5572, Observatoire Midi-Pyr\'en\'ees, 14 avenue E. Belin, 
F-31400 Toulouse, France}
\altaffiltext{5}{Present address: School of Physics and Astronomy, Tel Aviv University, 
69978 Tel Aviv, Israel; contini@wise.tau.ac.il}

\author{Daniel Kunth\altaffilmark{6}}
\altaffiltext{6}{Institut d'Astrophysique de Paris, 98 bis Bd. Arago, F-75014 Paris, France}

\and
\author{Georges Meynet\altaffilmark{3}}

\begin{abstract}
We present spectroscopic observations of the central star clusters in NGC 5253
the aim of which is to  search for WC stars.
Our observations show the presence of Wolf-Rayet (WR) stars not only of  
WN but also of WC subtype in two star forming regions corresponding to
the maximum optical and UV emission.
The massive star population we derive is consistent with young bursts of
$\sim$ 3 and 4 Myr.
The region of maximum optical emission is 
found to provide the dominant contribution of the ionizing flux 
as opposed to the less extinguished region of maximum UV brightness. 
The presence of WR stars near the N-enriched regions
found by Walsh \& Roy \markcite{wr87} \markcite{wr89} (1987, 1989) 
and Kobulnicky et al. \markcite{k96} (1997) suggests they are
a possible source of N.  
It is presently unclear whether or not our detection of WC stars is compatible
with the normal observed He/O and C/O abundance ratios.

\vspace*{0.5cm}
\centerline{\sl ApJ Letter, in press (Received: 2 November 1996,
Accepted: 13 March 1997)}
\end{abstract}


\twocolumn

\section{Introduction}
We have initiated  a systematic search program to find  Wolf-Rayet 
(WR) stars of the carbon series (WC subtypes) in the so-called WR galaxies
(Conti \markcite{c91} 1991).
Indeed, among the $\sim$ 70 WR galaxies known today only few cases
show the broad \civ\ $\lambda$5808 emission originating from WC stars 
(e.g.~Mrk 724, NGC~4861, NGC~4214, He~2-10, NGC~2363; cf.\ references
in Meynet \markcite{m95} 1995).
However, as Meynet \markcite{m95} (1995) pointed out, a natural result 
of stellar evolution (or our current knowledge of it) is that one expects that 
in a large fraction of WR galaxies 
(typically 30 \% for metallicities $1/5 \le Z/Z_\odot \le 1$; 
cf.~Schaerer \& Vacca \markcite{sv96a} 1996)  the WR population should be 
dominated by stars of the WC subtype if star formation 
occurs on timescales short compared to the lifetimes of massive stars.
Since the \civ\ $\lambda$5808 feature is usually weaker 
than the characteristic WR bump at $\sim$ 4700 \AA\  
it may simply have been overlooked in many previous observations.
Interestingly, in the homogeneous sample of low metallicity 
\hii\ galaxies of Izotov, Thuan \& Lipovetsky \markcite{i94}\markcite{i96}
(1994, 1997), 5 out of 15 
WR-rich objects show strong WC features, which corresponds surprisingly well 
to the theoretically predicted value of $\sim$ 30 \%.
Apart from the implications for the understanding of massive star evolution,
the presence or absence of a substantial population of WC stars in young
starbursts may imply significant differences of the ionizing spectrum
(cf.~Schaerer \markcite{s96} 1996) and the chemical enrichment due to massive stars
(cf.~Maeder \markcite{m83} 1983).

As one of our targets we have observed the central region of the 
amorphous galaxy NGC 5253 at a distance of 4.1 Mpc (Sandage et al. \markcite{s94} 1994). 
This low metallicity object ($Z/Z_\odot \sim 1/5$) contains a very young 
starburst and thus provides 
an important laboratory for studies of local chemical enrichment, or 
``chemical pollution'' in giant \hii\ regions (Walsh \& Roy \markcite{wr87}
\markcite{89} 1987, 1989, hereafter WR87, WR89; Pagel et al. \markcite{p92} 1992; 
Esteban \& Peimbert \markcite{ep95} 1995; Kobulnicky et al. \markcite{k96} 1997, 
hereafter KSRWR97).
In addition, recent HST imaging (Meurer et al. \markcite{m95} 1995; 
Gorjian \markcite{g96} 1996; Calzetti et al. \markcite{c97} 1997) 
reveals numerous Super Star Clusters in NGC 5253 and make this object
an interesting place to study what might be proto-globular clusters
(cf.~Conti \& Vacca \markcite{cv94} 1994; Meurer et al. \markcite{m95} 1995, 
and references therein).

In this Letter we report the detection of WR stars of both WN and WC 
subtype in two star forming regions of NGC 5253. 
The observations are presented in \S~2. In \S~3 and \S~4 we determine the 
massive star content of the two regions and ages of the populations. 
The implications regarding the source of ionization of NGC 5253,
the hardness of the ionizing flux, and the origin of N enrichment
are discussed in \S~5.

\section{Spectroscopic observations}
Long-slit spectra of NGC~5253 were obtained on the night of 1995 April 24 -- 
25 at the ESO\footnote{European Southern Observatory, La Silla, Chile} 2.2m 
telescope. The data were acquired with the EFOSC2 spectrograph and a 
1024$\times$1024 Thomson CCD with a pixel size of 0.34\arcsec. 
During the
photometric night we also observed the spectrophotometric standard stars 
HD~84937 and Kopff~27 in order to flux calibrate the spectra of the galaxy.
We used grism 4 which gives a spectral coverage of 4400 -- 6500 \AA\ with 
a resolution of $\sim$ 5 \AA.
The slit was oriented NNE-SSW (PA $\sim$ 20$^\circ$) in order to lie
the brightest optical regions of the galaxy.
The slit width was 1.6\arcsec\ for the galaxy observations 
and 5\arcsec\ for the standard stars. Spectra of He-Ar calibration lamp were 
obtained immediately before and after the galaxy integrations in order
to  accurately calibrate  the wavelength scale.
The total integration time of 5400 s was divided into 5 exposures (4 times 
20 minutes plus 10 minutes) in order to avoid saturation of the 
bright nebular lines (\hbeta\ and [\oiii]) and to recognize cosmic ray 
impacts. The seeing was relatively stable during the observations with a 
mean spatial resolution of about 1\arcsec. 
The spectrum was aquired at low airmass ($\sim$ 1.1) and no correction
for the loss of blue light due to atmospheric dispersion was made.

The spectra were reduced according to  standard reduction 
procedures using the MIDAS package LONG. These include bias subtraction, 
flat-field and airmass corrections, wavelength and flux calibrations, 
sky subtraction, and cosmic ray removal.
The spatial distribution of emission-lines along the slit reveals the 
presence of two bright regions, hereafter called {\em A} (for the brightest) 
and {\em B} (south of {\em A}), separated by about 3\arcsec\ (60 pc at a 
distance of 4.1 Mpc). 
Comparisons of our slit position with the HST images of Meurer et al. 
\markcite{m95} (1995, FOC with F220W filter) and Gorjian \markcite{g96} 
(1996, WFPC2 with F606W) show that 
region {\em A} is located at the maximum of the optical emission (=region 1 of
Gorjian), while {\em B} is centered on the maximum peak of UV emission
(=UV1 of KSRWR97). Note that region 1 of WR89 (=region A of WR87) contains 
both our regions {\em A} and {\em B}.
We also detected a  region (hereafter called {\em H}), located 
$\sim$ 3\arcsec\ NNE of {\em A}, for which no clear counterpart could be
found in the HST images.
We extracted one-dimensional spectra corresponding to the 
different regions by adding 3 columns ($\simeq$ 1\arcsec) along the 
spatial dimension.

\begin{figure*}[htb]
\centerline{\psfig{figure=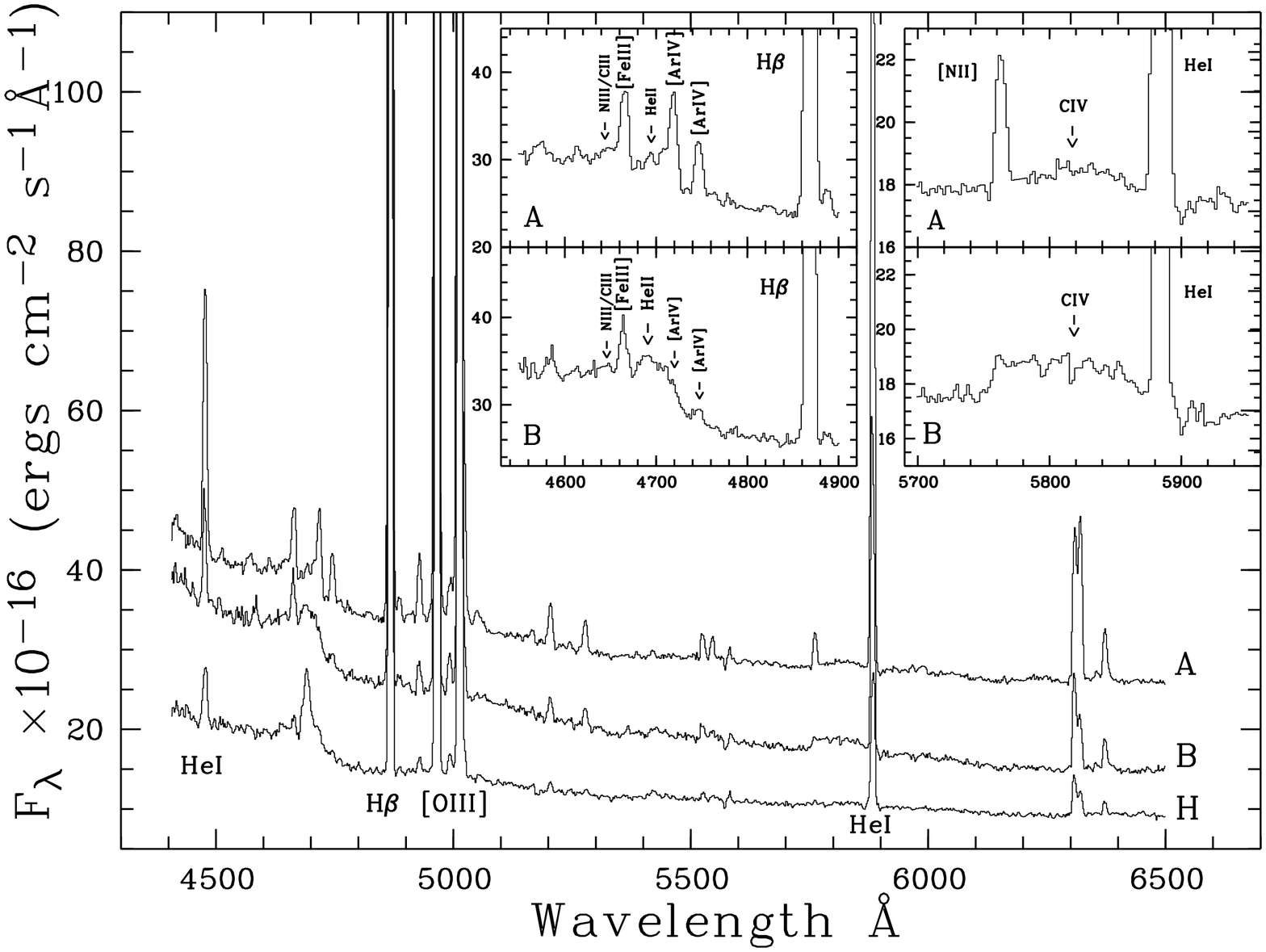,height=8.8cm,angle=270}}
\caption{Optical spectra of regions {\em A} (offset of +10 in 
flux units), {\em B} and {\em H} in NGC 5253. 
The insets show enlargements on the Wolf-Rayet bumps around 4700 \AA\ 
and 5800 \AA\  for regions {\em A} and {\em B}. 
The broad C~{\sc iv} $\lambda$5808 line from WC stars is detected in {\em A} 
and {\em B}. 
The broad He~{\sc ii} $\lambda$4686 line from Wolf-Rayet stars is clearly 
detected in {\em B} but is suspicious in {\em A} where the spectrum is 
dominated by high-excitation nebular emission lines of [Ar~{\sc iv}] and 
[Fe~{\sc iii}]. Note the bright nebular He~{\sc ii} line in the spectrum 
of region {\em H}. These spectra are not corrected for extinction
\label{fig_spectra}}
\end{figure*}

The two regions {\em A} and {\em B} are bright, 
low-metallicity \hii\ regions with a large number of intense nebular 
emission lines seen in their spectra. 
Because of the medium spectral resolution, the emission lines in the WR 
bumps at $\sim$ 4700 
and 5800 \AA\ (see Fig.~\ref{fig_spectra}) and the [\oi]/[\siii] lines at 
6300 \AA\ are not well separated. In order to measure accurately the 
individual emission 
lines, we use a procedure of multi-gaussians fitting where the number of 
gaussians is the only fixed parameter (position, intensity and FWHM
are free to vary).  
Line fluxes, FWHM and equivalent widths were determined using MIDAS 
standard commands. 
Due to the limited spectral range, our observations do not 
include the H$\alpha$ emission line commonly used to determine the 
internal reddening parameter $C_{\beta}$. 
For regions {\em A} and {\em H} we therefore adopt $C_{\beta} =$ 0.44
and 0.85 respectively, taken at the appropriate locations from the extinction
map of WR89. For {\em B} we adopt the reddening derived by KSRWR97 
($C_{\beta} =$ 0.20)\footnote{Note that highly spatially variable 
extinction has been suggested in NGC~5253 by KSRWR97 and Calzetti 
et al.\ \markcite{c97} (1997). IR and radio observations from Beck et al.\
\markcite{b96} (1996) show larger extinction than the optical data.}.
Dereddened fluxes are derived using the extinction law of Seaton
\markcite{s79} (1979) including a Galactic foreground extinction of  
E(B-V)=0.05 (Burstein \& Heiles \markcite{bh84} 1984). 
Observed and dereddened fluxes are given in Table~1.
The relative uncertainty of our measurements is smaller than 10\% for 
the most intense lines (\hbeta\ and [\oiii]), and can reach about 
40\% for the faintest. 
We note that our absolute flux calibration for region {\em A} agrees well 
with the values measured by Campbell, Terlevich \& Melnick \markcite{c86} 
(1986; hereafter CTM86) and WR89.


Up to three different broad emission lines (FWHM $\geq$ 30 \AA)
have been detected with 
varying confidence levels in regions {\em A} and/or {\em B}
over the wavelength range covered by our spectra:
the \niii/\ciii\ $\lambda\lambda$4640,4650 blend, \heii\ $\lambda$4686,
and \civ\ $\lambda$5808-5812.
Clear detections ($\geq$ 3 $\sigma$) of broad lines are
\civ\ in {\em A} and {\em B} and \heii\ $\lambda$4686 in {\em B}.
The detection and hence the FWHM of \heii\ $\lambda$4686 in the spectrum 
of {\em A} dominated by high-excitation nebular emission lines of 
[\ariv] $\lambda\lambda$4711,4740 (see Fig.~\ref{fig_spectra})
is uncertain ($\sim 2 \sigma$).
\niii/\ciii $\lambda\lambda$4640,4650 
may be present in {\em A} and {\em B} at a 1-2 $\sigma$ level.
Region {\em H} stands out with a very large \heii\ $\lambda$4686/\hbeta\
ratio, uncommon for the metallicity of NGC 5253, and a relatively 
narrow \heii\ $\lambda$4686 line (FWHM=20 \AA). Due to these peculiarities and
the absence of a visible counterpart in any images (cf.\ above) this 
region is not discussed further. Future studies would be highly desirable.

The broad WR bump around 4700 \AA\ has been previously detected  by CTM86
\markcite{c86}, 
\markcite{wr87} WR87 in their region 1 (containing both our regions {\em A} and {\em B}), 
and in the region UV1 (=our $B$ region) of \markcite{k96} KSRWR97.
One of the main results of this Letter is the unambiguous detection 
of a broad (FWHM = 62 and 86 \AA) \civ\ $\lambda$5808 emission line in 
regions {\em A} and {\em B}  (see Fig.~\ref{fig_spectra}), which 
clearly indicates the presence of WC stars in these regions (see \S~\ref{s_wrlines}). 
A sufficiently high S/N is required for the detection of the \civ\ $\lambda$
5808 line in integrated spectra of extragalactic \hii\ regions or alike objects,
and to derive the observed frequency of WC stars in WR galaxies.
This explains the non detection of this line in previous ground-based spectra 
of NGC 5253 (CTM86, WR87, WR89) \markcite{c86} \markcite{wr87} \markcite{wr89}
and in the recent HST FOS spectrum of the region UV1 
of \markcite{k96} KSRWR97.
Interestingly, however, the presence of WC stars was suspected by WR89 
\markcite{wr89} based on a high carbon abundance derived from IUE spectra 
(cf.~\S 4).

\section{The origin of the broad emission lines in regions {\em A}
and {\em B}}
\label{s_wrlines}
Both from the relative strength of \heii\ $\lambda$4686 to \civ\ $\lambda$5808
and from the large FWHM of the latter line,
WN stars can be ruled out as the origin of the \civ\ line. 
The observed FWHM ($\sim$ 70 \AA) of this line corresponds well to early type WC stars 
(typically WC4; Smith, Shara \& Moffat \markcite{ssm90b} 1990) including 
possibly also some WO3-4 stars:
their \ion{O}{5} $\lambda$5590 emission ($<$ 15 \% of \civ\ $\lambda$5808) 
would remain undetectable. From the absence of \ion{O}{5} $\lambda$5590 a 
significant number of WO1-2 stars can 
be excluded (cf.~Kingsburgh, Barlow \& Storey \markcite{k95} 1995).
Similarly the absence of \ciii\ $\lambda$5696 excludes a significant
population of late type WC stars.
The relative contributions of the \niii/\ciii\ $\lambda\lambda$4640,4650 
blend and \heii\ $\lambda$4686 to the WR bump exclude a pure WC origin.
The FWHM ($\sim$ 30 \AA) of \heii\ $\lambda$4686 agrees well with WNL stars
(Smith, Shara \& Moffat \markcite{ssm96} 1996). 
All broad emission line features therefore indicate a mixed
population of WNL, and WC4 and/or WO3-4 stars in regions {\em A} and {\em B}.
This result is further supported by quantitative comparisons
presented in \S~4.
The presence of intermediate type WN/WC stars cannot be excluded.

\section{The Wolf-Rayet and O star population in NGC 5253}
The approximate number of O and WR stars present in the observed regions 
can be estimated as follows.
Adopting the value $\log Q_0=49.05$ from Vacca \& Conti \markcite{vc92} 
(1992, hereafter VC92) for the Lyman continuum luminosity of a O7V star 
and assuming Case B 
recombination, we estimate  $\sim$ 865 (210) O7V stars 
in region {\em A} and in region {\em B}  from their \hbeta\ luminosity.
These numbers are approximately $\sim$ 5 and $\sim$ 35 \% lower,
respectively, if we account for the contribution of the WR stars to the 
ionizing flux.
Depending on the age of the population and the IMF slope, the number of 
O7V stars does not necessarily correspond well to the total number of O 
stars: most likely the total number of O stars is larger 
by a factor of $\sim$ 2 and 4 for {\em A} and {\em B} respectively 
(Schaerer \markcite{s96} 1996).
Using the average observed luminosity of WC4 stars in the 
\civ\ $\lambda$5808 line ($L_{5808} = 2.5 \times 10^{36} {\rm erg \, s^{-1}}$;
VC92), the 5808 emission in 
regions A and B can be explained by 10 and 13 WC4 stars respectively. 
From the observed \heii\ $\lambda$4686 line luminosity we derive an 
upper limit of 25 WNL stars for {\em A} (depending on nebular
contribution) and 39 WNL stars for {\em B}, using the average 
line luminosity given by \markcite{vc92} VC92.

\begin{figure}[htb]
\centerline{\psfig{figure=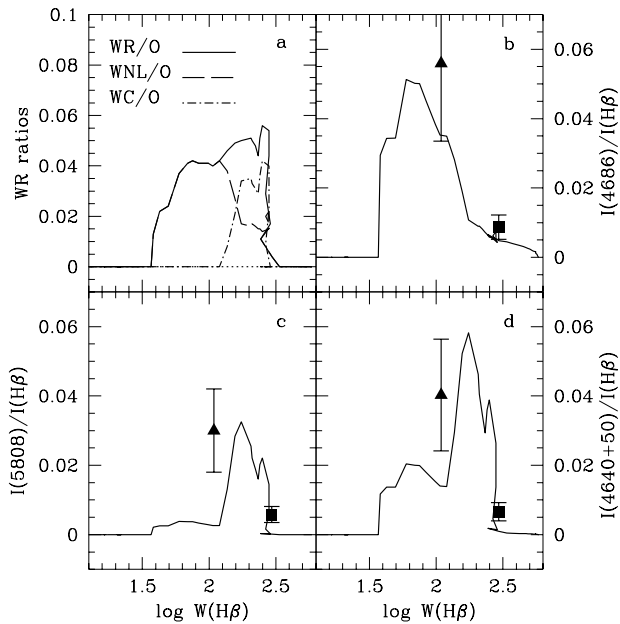,height=8.8cm}}
\caption{
The predicted evolution of massive star 
populations and emission-line 
ratios from evolutionary synthesis models as a function of the \hbeta\
equivalent width. All models are given for Z=0.004, assuming an 
instantaneous burst and a Salpeter IMF with $M_{\rm up}=$ 120 \msun.
Observations for regions A and B are shown by squares and triangles
respectively in panels b-d. The error bars correspond to the estimated 
uncertainty of $\pm$ 40 \% typical for weak lines.
{\bf a.} Total WR/O number ratio (solid), WNL/O (dashed), and WC/O ratio
(dashed-dotted).
{\bf b.} Ratio of the predicted (He~{\sc ii} 4686/H$_\beta$) flux.
{\bf c.} Same as b for C~{\sc iv} 5808.
{\bf d.} Same as b for the N~{\sc iii}+C~{\sc iv} blend at 4640-4650
\label{fig_models}}
\end{figure}

To see whether the WNL, WC and O star populations are compatible
with predictions from stellar evolution we use the recent evolutionary
synthesis models of Schaerer \markcite{s96} (1996) and Schaerer \& Vacca 
\markcite{sv96a} (1996), which
allow direct comparisons to be made with the relevant observational quantities 
related to the WR and O star population (Fig.~\ref{fig_models} panels b-d).
The predictions are shown for instantaneous burst models at the 
appropriate metallicity for regions {\em A} and {\em B} (Z=0.004, 
cf. \markcite{wr89} WR89, \markcite{k96} KSRWR97), assuming a Salpeter IMF 
with an upper mass cut-off of $M_{\rm up}=$ 120 \msun.
The observed line fluxes in the different WR lines (in units of the 
\hbeta\ flux) are shown as filled symbols.
Panel {\bf a} illustrates the corresponding WR/O, WNL/O, and WC/O number ratios.
Figure \ref{fig_models} shows that both the observed values of the WR lines
fluxes and their variation with $W(\hbeta)$ can be reasonably well 
reproduced by the models for Z=0.004.
With several variations of the input parameters (flatter IMF, short burst 
duration, etc.) all the observed values can be matched with greater accuracy.
In view of the uncertainties of the measurement for these weak lines 
we conclude \hbeta\ and all broad WR lines in both regions {\em A} and
{\em B} are consistent with an approximately instantaneous burst.
The strong WR features also imply the presence of very massive stars
($M_{\rm initial} \ge$ 60 \msun).

In the burst models a  decrease of $W(\hbeta)$ corresponds to an increasing 
age of the burst. The ages derived from our models correspond to $\sim$ 
2.8 Myr for {\em A} and 4.4 Myr for {\em B}. As shown above this age
sequence is also compatible with the WR line fluxes, which exhibit 
changes on short time scales.
The non-detection of radio emission from SNR in NGC~5253
supports the presence of very young nuclear starbursts
(less than 1-2 10$^7$ yr, Beck et al. \markcite{b96} 1996).

\section{Discussion}
Does the presence of WC stars lead to the  high excitation of the gas,
 and thus the nebular
\heii\ emission as suggested by Schaerer \markcite{s96} (1996) ?
At the age of region {\em A} (estimated from its large value of 
W(\hbeta)) the models of Schaerer \markcite{s96} (1996) predict an important 
{\em nebular} contribution to the total \heii\ $\lambda$4686 emission due 
to the large fraction of WC stars in the burst.
Given the weakness of the 4686 line in this region, its width is quite
uncertain.
The presence of strong nebular lines of [\feiii] $\lambda$4658 and [\ariv] 
$\lambda\lambda$4711,4740 indicating high excitation
resembles closely the cases of Pox~4 (cf.~Kunth \& Sargent \markcite{ks81} 
1981; \markcite{vc92} VC92) and Mkr~1271 (Contini \markcite{c96}
1996) where the distinction between broad stellar and nebular \heii\ is unclear. 
The observations of region {\em A} are thus compatible with WC stars as the 
origin of nebular \heii\ emission.
At the age corresponding to region {\em B} the models of Schaerer \markcite{s96}
(1996) predict only  broad \heii\ emission, in agreement with the observations.
Compared to {\em A}, the weakness of the forbidden \ion{Ar}{4} lines also 
indicates a lower excitation.

The analysis of the stellar population in regions {\em A} and {\em B} sheds
further light on the ionizing source of NGC 5253 and has important implications
for scenarios of local chemical enrichment.

{\em The ionizing clusters:}
As pointed out by KSR\-WR\-97 the brightest UV cluster (their UV1 which is 
included in our region {\em B}) does not provide enough ionizing flux to 
explain the \halpha\ surface brightness of NGC 5253 corresponding
to $\sim 4 \times 10^{52}$ photons s$^{-1}$ (Martin \& Kennicutt \markcite{mk95}
1995) derived assuming a constant extinction of $C_{\beta} =$ 0.47.
Our observations, however, clearly reveal that region {\em A} 
(corresponding to the optical maximum and also approximately to the \halpha\ 
emission peak) contains a larger number of O stars than region {\em B}.
From \hbeta\ we obtain an ionizing Lyman continuum flux of
$9.7 \times 10^{51}$ and $2.4 \times 10^{51}$
photon s$^{-1}$ for regions {\em A} and {\em B} respectively.
Region {\em A} is of higher extinction and thus clearly dominates 
the production of ionizing photons as compared to the brightest region in 
the UV. 
In view of the possible underestimation of extinction
(cf.~Beck et al. \markcite{b96} 1996) and the strong spatial
extinction variations 
(KSRWR97, Calzetti et al. \markcite{c97} 1997) we suspect that the close area 
surrounding region {\em A} (possibly including also {\em H}) may well produce enough 
ionizing photons to explain the total observed \halpha\ surface brightness.

{\em WR stars and the chemical enrichment:}
Local overabundance of nitrogen in several regions of NGC 5253 is well
established \markcite{wr87} \markcite{wr89} \markcite{k96} (WR87, WR89, KSRWR97).
The recent HST observations of KSRWR97 show N enrichment in two locations
(called HII-1 and HII-2) close to the peak of \halpha\ emission (= our
region {\em A}), while N/O in their region UV1 (included in our region
 {\em B}) seems to be typical for metal-poor galaxies.
Assuming that WR stars eject a significant amount of nitrogen, our 
likely detection of WR stars in {\em A} provides a plausible source of 
N in the very proximity of the N-enriched regions.
Region {\em A}, overlooked by KSRWR97 due to its low UV brightness, is thus 
most likely the ``hidden'' cluster containing the  sources of the
 observed pollution.

More generally, the detection of WR stars of both WN and WC subtypes 
in both of our regions raises several questions. 
The earlier finding of a possible carbon overabundance by \markcite{wr89}
WR89, which originally lead these authors to suggest the presence of WC stars, 
was not confirmed by the results of \markcite{k96} KSRWR97.
If the overabundance of N (in the regions close to {\em A}) is due to 
WN stars, why are the ejecta of WC stars (presumably mostly He, C and O) 
not detected ?
Furthermore, why does region {\em B} (=UV1), which also harbours WR stars, 
not show any significant overabundance ?
Abundance differences between these two regions might be related to their 
age difference: maybe mixing processes in {\em B} had time to dilute 
the ejecta with the ambient medium, whereas the younger region {\em A} 
is currently in a phase of heavy pollution.
Hopefully a study of NGC 5253 will not only tell us its history of 
star formation but also more about still poorly understood 
nucleosynthetic yields in massive stars and mixing processes in the ISM.

\acknowledgments
DS thanks Daniela Calzetti, Rosa Gonz\'{a}lez-Delgado, Claus Leitherer,
Gerhard Meurer, and Bill Vacca for stimulating and useful discussions.
DK thanks Eline Tolstoy for a careful reading of the manuscript.
Daniela Calzetti also kindly provided us with HST images of NGC 5253.
DS acknowledges a fellowship from the Swiss National Foundation 
of Scientific Research and partial support from the Directors 
Discretionary Research Fund of the STScI.


\clearpage

\clearpage

\newpage
\begin{deluxetable}{lrrrrrrr}
\scriptsize
\tablenum{1}
\tablewidth{0pt}
\tablecaption{Normalized Flux of Emission Lines in Three Knots of NGC 5253 \label{table1}}
\tablehead{
\colhead{Line}             & 
\colhead{$\lambda$}        &
\multicolumn{6}{c}{Region} \nl
\colhead{}                 &
\colhead{(\AA)}            &
\multicolumn{2}{c}{A}      &
\multicolumn{2}{c}{B}      &
\multicolumn{2}{c}{H}     \nl
\colhead{}                 &
\colhead{}                 &
\colhead{$F_{\lambda}$}    &
\colhead{$I_{\lambda}$}    &
\colhead{$F_{\lambda}$}    &
\colhead{$I_{\lambda}$}    &
\colhead{$F_{\lambda}$}    &
\colhead{$I_{\lambda}$}
}               
\startdata
He\,{\sc i} & 4471 & 45.8 & 52.1 & 36.0 & 38.4 & 55.2 & 69.8 \nl
N\,{\sc iii}/C\,{\sc iii}\tablenotemark{a} & 4645 & (6.2) & (6.7) & (39.3) & 
(40.7) & (35.4) & (40.1) \nl
[Fe\,{\sc iii}] & 4658 & 13.0 & 13.8 & 19.1 & 19.8 & 22.7 & 25.5 \nl
He\,{\sc ii}\tablenotemark{a} & 4686 & 8.3 & 8.8 & 54.8 & 56.3 & 150.3 & 166.3 \nl  
[Ar\,{\sc iv}] & 4711 & 16.1 & 16.9 & (21.9) & (22.3) & (48.3) & (52.6) \nl 
[Ar\,{\sc iv}] & 4740 & 10.1 & 10.5 & (10.8) & (11.1) & (12.7) & (13.6) \nl 
He\,{\sc i} & 4922 & 10.0 & 9.8 & 10.9 & 10.7 & 10.1 & 9.7 \nl 
[O\,{\sc iii}] & 4959 & 2123.0 & 2059.7 & 1564.1 & 1538.8 & 1779.4 & 1685.1 \nl 
[O\,{\sc iii}] & 5007 & 6375.6 & 6103.1 & 4695.2 & 4589.0 & 5322.4 & 4917.1 \nl 
[N\,{\sc i}] & 5199 & 7.4 & 6.7 & 7.4 & 7.0 & \nodata & \nodata \nl 
[Fe\,{\sc iii}] & 5271 & 4.8 & 4.2 & 6.4 & 6.0 & \nodata & \nodata \nl
[Cl\,{\sc iii}] & 5518 & 4.0 & 3.3 & 4.6 & 4.1 & \nodata & \nodata \nl 
[Cl\,{\sc iii}] & 5538 & 2.9 & 2.4 & \nodata & \nodata & \nodata & \nodata \nl 
[N\,{\sc ii}] & 5755 & 5.4 & 4.3 & (2.8) & (2.4) & \nodata & \nodata \nl
C\,{\sc iv}\tablenotemark{a} & 5808 & 7.2 & 5.6 & 33.0 & 29.0 & \nodata & \nodata \nl
He\,{\sc i} & 5876 & 142.2 & 109.5 & 127.6 & 111.0 & 126.9 & 79.2 \nl  
[O\,{\sc i}] & 6300 & 26.8 & 19.0 & 34.7 & 28.8 & 33.5 & 17.9 \nl  
[S\,{\sc iii}] & 6312 & 40.2 & 28.4 & 24.7 & 20.6 & 22.1 & 11.8 \nl  
[O\,{\sc i}] & 6364 & 8.8 & 6.2 & 10.5 & 8.7 & 10.0 & 5.2 \nl  
\hline
 & & & & & & & \nl
$C_{\beta}$ & & 0.44 & & 0.20 & & 0.85 & \nl
$W(H\beta)$& & 294 & & 109 & & 87 & \nl
$F(H\beta)$ & & 70.6 & 229.8 & 29.8 & 55.8 & 13.1 & 109.6 \nl
\tablenotetext{a}{Broad emission lines (FWHM $\geq$ 20 \AA) from 
Wolf-Rayet stars.}
\tablecomments{For each emission-line we reported the observed 
($F_{\lambda}$) and dereddened ($I_{\lambda}$) flux normalized to 
$H\beta \times 1000$. Values in brackets are uncertains. 
$C_{\beta}$ is the extinction coefficient, $W(H\beta)$ is the equivalent width 
of H$\beta$ in \AA\ and $F(H\beta)$ are the absolute observed and dereddened 
flux of H$\beta$ emission-line in $10^{-14}$ erg cm$^{-2}$ s$^{-1}$. 
}
\enddata
\end{deluxetable}

\end{document}